\documentclass{mn2e}
\usepackage{psfig}
\usepackage[authoryear]{natbib}

\def \etal{\rm et~al.~}
\def \saxx{\rm 1SAX~J1218.9+2958}
\newcommand{\nh}{{\rm N_{H}}}

\newcommand{\rosat}{{\it ROSAT}}
\newcommand{\asca}{{\it ASCA}}
\newcommand{\chandra}{{\it Chandra}}
\newcommand{\xmm}{{\it XMM--Newton}}
\newcommand{\sax}{{\it Beppo}SAX}

\newcommand{\aox} {{$\alpha_{\rm ox}$}}

\newcommand{\lsim}{{\lower.5ex\hbox{$\; \buildrel < \over \sim 
\;$}}}
\newcommand{\gsim}{{\lower.5ex\hbox{$\; \buildrel > \over \sim 
\;$}}}

\title[\saxx]
      {The obscured QSO 1SAX J1218.9+2958}
\author[N.S. Loaring $\etal$]{N.S. Loaring$^{1}$ 
\thanks{nsl@mssl.ucl.ac.uk}, M.J. Page$^1$, G. Ramsay$^1$\\
{$^1$Mullard Space Science Laboratory, 
University College London, Dorking, Surrey, RH5 6NT, UK}\\ }

\begin{document}

\maketitle

\begin{abstract}
We present results from \xmm\ observations of the obscured QSO \saxx.
We find that the previously reported optical and soft X--ray counterpart 
positions are incorrect. However we confirm the spectroscopic redshift of 
0.176.
The optical counterpart has a K magnitude of 13.5 and an 
R--K colour of 5.0 and is therefore a bright extremely
red object (ERO).
The X--ray spectrum is well  described by a power-law
($\Gamma=2.0\pm0.2$) absorbed by 
an intrinsic neutral column
density of $8.2^{+1.1}_{-0.7}\times10^{22} \rm{~cm^{-2}}$.
We find that any scattered emission contributes at most 0.5 percent to the
total X--ray flux. 
From the optical/near--IR colour 
we estimate that the active nucleus must contribute at least 50 percent of the
total flux in the K band and that the ratio of extinction to X--ray absorption is 
$0.1-0.7$ times that expected
from a Galactic dust-gas ratio and extinction curve.
If \saxx\ were $100\times$ less luminous it would be indistinguishable from the
population responsible for most of the 2--10 keV X--ray
background. 
This has important implications for the optical/IR properties of faint 
absorbed X--ray sources. 

\end{abstract}


\section{Introduction}

The cosmic X--ray background (XRB) is the combined X--ray emission from all
extragalactic X--ray source populations, integrated over cosmic time.  Surveys
made using \rosat\ resolved 70--80 percent of the XRB 
into discrete sources \citep{hasinger98} at soft energies (0.5--2.0 keV). 
The majority of
these sources are  QSOs and Seyfert 1 galaxies: unobscured active
galactic nuclei (AGN) with broad emission lines
\citep{mchardy98,lehmann01}. 
However, the XRB cannot be explained solely by extrapolating these sources to
fainter fluxes as their spectra are softer than that of the XRB. To reproduce
the XRB spectrum, an additional population of faint hard sources is required.
Leading X--ray background synthesis models identify
the faint hard sources with intrinsically absorbed AGN 
\citep{comastri95,wil00,gilli01}. The absorbed AGN
are expected to constitute 80 to 90 percent of 
the total AGN population \citep{fabian99}.

The anticipated population of faint, hard sources has now been detected in
\chandra\ and \xmm\ surveys \citep{tozzi01,brandt01,hasinger01}.  
However, even before the launch of \xmm\ and \chandra, a
number of X--ray surveys were carried out using \rosat, \asca, and \sax\ to
investigate the absorbed AGN population \citep{ueda99,fiore99,page00}. 
These surveys covered large areas of sky and so were particularly useful for
finding relatively bright, and thus easily studied, examples of the absorbed
AGN population.

The sources detected in these surveys exhibit interesting, and somewhat
surprising properties. 
For example, \citet{maio00}
found that AGN from the \sax\ High Energy Large Area Survey (HELLAS) 
could be divided
into two groups according to their optical and near--IR properties. 
One group
of sources have point-like optical/near--IR counterparts with colours
typical of optically selected QSOs. The other group of sources are extended in
the optical/near--IR and have colours dominated by the host galaxy stellar
population. Despite having a wide range of X--ray absorbing column densities
up to $10^{23}~{\rm cm^{-2}}$, none of the sources 
studied by \citet{maio00} appeared to
have colours dominated by a reddened AGN.

Comparison of \rosat\ and \sax\ data on HELLAS sources yielded another
unexpected result: \citet{vignali01} found that a
significant fraction of the absorbed objects in HELLAS have an additional soft
component in their X--ray spectra. This could be due to scattered X--rays which
do not pass through the absorber, alternatively it might indicate a hot gas
component as observed in starburst galaxies.

Perhaps one of the most puzzling results is that in many AGN whose X--ray
spectra indicate a substantial absorbing column ($>10^{22}$ cm$^{-2}$), the
corresponding optical spectrum shows little or no evidence for dust reddening
\citep{akiyama00,page01,comastri01}.  This suggests that the absorbing
media of AGN typically have much lower dust-gas ratios than the interstellar
medium of our Galaxy \citep{maio01a}.

Observations of hard X--ray selected, absorbed AGN have the potential to reveal
much about the dominant, absorbed XRB-producing population of AGN. Here  we 
report new X--ray and optical observations of the HELLAS
source \saxx, an absorbed X--ray source, optically classified as a
Seyfert 1.9 galaxy \citep{fiore99}. It has a broad component to its H$\alpha$ emission line (hence
the Seyfert 1.9 classification) but its optical/near--IR colours appeared to 
be completely dominated by its host galaxy \citep{maio00}. 
It was proposed to have a \rosat\ counterpart, (and thus a significant soft
X--ray excess) by \citet{vignali01}.
 
Throughout this paper we assume a flat $\Lambda CDM$ cosmology with $(\Omega_M, \Omega_\Lambda)=(0.3,0.7)$ and a Hubble constant of
70 $\rm{km~s^{-1}~Mpc^{-1}}$.

\section{Observations and data reduction}
\subsection{\xmm\ data}
   
\saxx\ is a serendipitous source in two \xmm\ observations of Markarian
766. The first observation was performed on 20th May 2000 and had an exposure
time of 28~ks in EPIC MOS \citep[an intermediate resolution X--ray imaging
spectrometer,][]{turner2001}. 

The second observation took place exactly one year
later, with a MOS exposure time of 127~ks. In both observations the EPIC PN 
camera was operated in small window mode, and hence did not record any data 
for \saxx.

The data were processed using the \xmm\ Science Analysis System (SAS) v5.3.
Periods of high background in the 2001 observation were excised from the event
lists, to leave 95 ks of useful exposure time.
There were no periods of high background during the May 2000 observation, so
the total useful exposure time of the two observations is 123 ks. At an off
axis angle of 8.6 arc minutes, this corresponds to an effective exposure time
of 82 ks  at 2 keV.

For the purpose of obtaining an accurate source position, images and exposure
maps were constructed in the energy
band 1--10 keV. These images were source searched using the SAS routines
EBOXDETECT and EMLDETECT. A linear shift of $\sim$ 2 arc seconds, based on the
relative position of Markarian~766 in the EPIC images and the 2MASS catalogue, 
was applied 
to the X--ray position of \saxx\ to
match the optical/IR astrometric reference frame.

We extracted X--ray spectra in the energy range 0.2--12 keV for the source
and background from each exposure, including all valid single, double and
triple X--ray event patterns. We have used the latest EPIC MOS 
redistribution matrix files and generated appropriate ancillary response files
for our spectra using the SAS task ARFGEN. The source has a similar countrate
and similar X--ray spectrum in both observations. Therefore the spectra from
both observations and both MOS cameras were co-added 
to maximise signal to noise.
Finally, the resulting spectrum was grouped to a minimum of 50
counts per channel.

\subsection{Optical/IR data}

The online APM\footnote{http://www.ast.cam.ac.uk/$\sim$mike/apmcat/} and
2MASS\footnote{http://www.ipac.caltech.edu/2mass/} databases were used to
find the 
optical/IR 
counterpart for \saxx. In addition, an I band image was obtained
in January 2002 during photometric conditions  using the Andalucia Faint Object
Spectrograph and Camera (ALFOSC) on the Nordic Optical Telescope (NOT). 
An optical spectrum of \saxx\ was also obtained using ALFOSC in January
2002. A ten minute integration was taken using Grism 4 (3200--9100 \AA) with a
slit 1.8 arc sec wide providing a resolution of 23 \AA.
Relative flux calibration was determined from an observation of the standard
star Feige~66.

\section{The X--ray position of $\saxx$ and identifying the Optical/IR 
counterpart}

The \xmm\ derived position for \saxx\ is given in 
Table \ref{tab:positions}. 
The statistical uncertainty on this position is sub-arc second, and the
systematic uncertainty is $\sim1$ arc sec \citep{watson}, a significant
improvement upon the $\sim1$ arc minute uncertainty of the HELLAS 
position \citep{fiore99}. 

\begin{figure}
\begin{center}
\leavevmode
\psfig{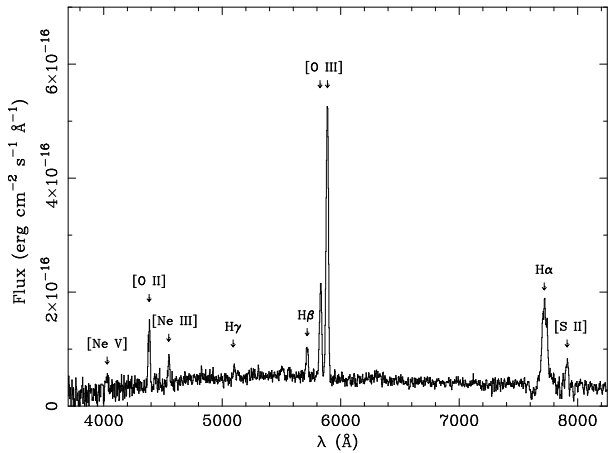}
\caption{Optical spectrum of \saxx. Prominent emission lines are
marked for $z=0.176$.}  
\label{fig:optspec}
\end{center}
\end{figure}

At the X--ray position we find 
unambiguous counterparts in the 
APM and 2MASS
catalogues, and their positions are given in Table \ref{tab:positions}.
There is no corresponding \rosat\ counterpart.
We provide a summary of the optical/IR counterpart's photometric 
properties in Table 2. 
The spectrum of the optical counterpart is shown in Figure
\ref{fig:optspec}. It is a Seyfert 1.9 galaxy at a redshift
$z=0.176$. 
The position of the optical counterpart
reported by the HELLAS team \citep{fiore99,maio00,lafranca02} is
incorrect. 
However, the spectrum of the optical counterpart 
shown in \cite{fiore99} has exactly the same emission line ratios as seen in
our NOT spectrum. Subsequently we have learned that the spectrum and redshift
given by \citet{fiore99} are in fact for the correct optical counterpart, but
the reported co-ordinates were mistakenly given for a second object observed 
simultaneously through the same slit (A. Comastri, private communication). 

\begin{table}
\baselineskip=20pt
\begin{center}
\begin{tabular}{lcc}  
& \multicolumn{2}{c}{RA (J2000) Dec}  \\
&&\\
\xmm & 12:18:54.73 & +29:58:06.6 \\
APM & 12:18:54.71 & +29:58:06.7 \\
2MASS &12:18:54.68 & +29:58:06.7\\
\end{tabular}
\end{center}
\caption{X--ray, optical and IR positions of $\saxx$}
\label{tab:positions}
\end{table}

\begin{table*}
\baselineskip=20pt
\begin{center}
\begin{tabular}{cccccccc}
\hline
\multicolumn{2}{c}{2MASS} &POSS &POSS & NOT &\multicolumn{3}{c}{2MASS}   \\  
\multicolumn{2}{c}{RA (J2000) Dec} & O (B) & E (R) &  I & J & H  & K \\  
\hline   
12 18 54.68 & +29 58 06.7 & 20.07 (19.88) & 18.52 (18.52) & 17.95$\pm0.1$ & 16.425 & 14.935 & 13.499 \\
\hline 
\end{tabular}
\caption{Optical and IR colours of the source.}
\end{center}
\label{tab:opt_mags}
\end{table*}

\section{X--ray spectral analysis}
\label{spectrum}

We have used {\small XSPEC}~v11.2 to analyze the X--ray spectrum. The results 
described in this section are given in Table 
\ref{tab:results} and all errors on the fit
parameters are quoted at 90 per cent confidence for one interesting parameter
($\Delta\chi^{2}=2.706$).

\begin{figure}
\setlength{\unitlength}{1in}
\begin{picture}(2.9,2.5)
\put(0.0,3.0){\includegraphics{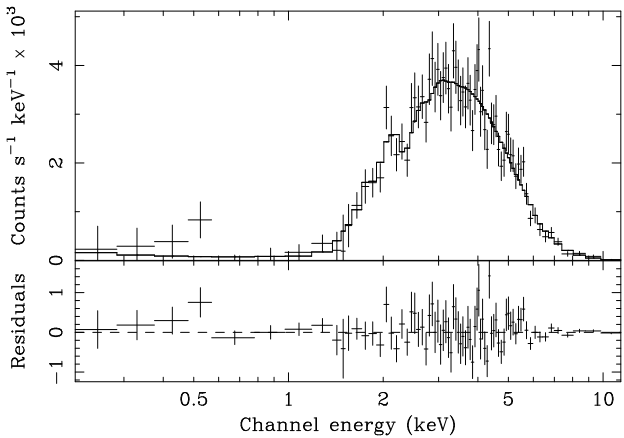}}
\end{picture}
\caption{The top panel shows the observed EPIC MOS spectrum of \saxx\ 
(datapoints) fitted with an absorbed power 
law model (histogram). The bottom panel shows the residuals to the model fit.}
\label{fig:xmmspec}
\end{figure}

\subsection{Absorbed power-law model}

The spectrum is shown in Figure \ref{fig:xmmspec}; the low count rate below 
1 keV implies that the spectrum is heavily absorbed. 
We therefore fitted an absorbed  power-law model. 
A Galactic component to the absorption
model was included, with $\nh = 1.7\times 10^{20}{\rm~cm^{-2}}$, \citep{dickey90}.
This simple model is a good fit to the data and
is overplotted in Figure \ref{fig:xmmspec}. 
The
best fit power-law photon
index is $\Gamma=2.0\pm0.2$, and the best fit column
density intrinsic to \saxx\ is 
$8.2^{+1.1}_{-0.7}\times 10^{22}{\rm ~cm^{-2}}$. 
The observed 2--10 keV flux is $8.6\times10^{-13}\rm{~erg~s^{-1}~cm^{-2}}$,
with a corresponding unobscured rest frame 2--10 keV luminosity of 
$1.1\pm0.1\times 10^{44}{\rm~erg~s^{-1}}$.

\subsection{Reflection and reprocessing}

Reprocessing of the primary power-law radiation by cold material will result in
a Fe $K_{\alpha}$ emission line at 6.4 keV and a reflected component, which has
an Fe K edge at 7.1 keV  \citep{george91,matt91}. Addition of a narrow line 
at a rest frame energy of 6.4 keV does not produce a significant  improvement 
to the fit. Adding a cold reflection component also results in an 
insignificant improvement in fit. The reflection component is poorly 
constrained and the 90 
percent upper limit to the equivalent width of the Fe $K_\alpha$ line (111 eV)
is compatible with the typical level of reflection seen in Seyfert 1 galaxies
\citep{nandra97} and with the reprocessing expected in the absorbing 
medium \citep{leahy93}. 

Some Seyfert 2 galaxies show broad optical emission lines in
polarised light \citep[e.g.][]{am85}, implying that some of the
nuclear radiation is scattered into our line of sight without passing through
the absorbing medium. 
We have therefore added a second intrinsically unabsorbed 
power-law component to our model. 
This does not significantly improve the fit, and the  
normalisation of the scattered power-law component is 
$<0.5\%$ of the primary, absorbed, power-law component.
In several heavily absorbed Seyfert galaxies the spectrum below 1 keV is
dominated by emission from photo-ionised gas \citep{kinkhabwala02,sako00}.
The strongest soft X--ray emission feature in these objects is
the OVII triplet of emission lines at rest frame energy of 0.57 keV which is
unresolved at the EPIC spectral resolution. Addition of a  narrow rest frame 
0.57 keV emission line does not significantly improve the fit. 

\begin{table*}
\baselineskip=25pt
\begin{center}
\begin{tabular}{lccccccc}
\hline
Model  & $\Gamma$ & Intrinsic $N_{H}$ & EW & Line flux & Scattered  &
R  & $\rm{\chi^{2}/d.o.f.}$ \\
& & ($10^{22} {\rm ~cm^{-2}}$) & (eV) & ($\rm{10^{-14}~erg~cm^{-2}~s^{-1}}$) &
fraction (\%)\\
\hline
Absorbed power-law & $2.0^{+0.2}_{-0.2}$ & $8.2^{+1.1}_{-0.7}$& & &
& & 69/73\\
\\
Absorbed power-law & $2.0^{+0.3}_{-0.2}$ &
$8.3^{+1.1}_{-0.7}$ & $56^{+57}_{-53}$ & $0.8^{+0.8}_{-0.8}$  &  & & 66/72\\
+ iron $K_{\alpha}$ line \\
Absorbed power-law & $2.3^{+0.3}_{-0.4}$
& $8.9^{+0.9}_{-0.8}$ & $52\pm52$ & $0.7^{+0.8}_{-0.7}$ & & $1.8^{+6.0}_{-1.8}$ & 66/71
\\
+ reflection + iron $K_{\alpha}$ line \\
Absorbed power law & $2.0^{+0.2}_{-0.2}$ &
$8.4^{+1.1}_{-0.9}$ & & & $0.2^{+0.3}_{-0.2}$  & & 68/72\\
+ scattered component \\
Absorbed power-law & $2.0^{+0.1}_{-0.1}$ &
$8.3^{+0.8}_{-0.9}$ & - & $0.1^{+0.1}_{-0.1}$ & & & 66/72\\
+ OVII line \\
\hline
\end{tabular}
\end{center}
\caption{X--ray spectral model fitting results. R is the normalisation of the
reflection component compared to a plane reflector covering half of the sky as
seen from the central source.}
\label{tab:results}
\end{table*}

\section{AGN contribution to the optical/IR flux}

\begin{figure}
\setlength{\unitlength}{1in}
\begin{picture}(2.9,2.5)
\put(0.0,-2.2){\includegraphics{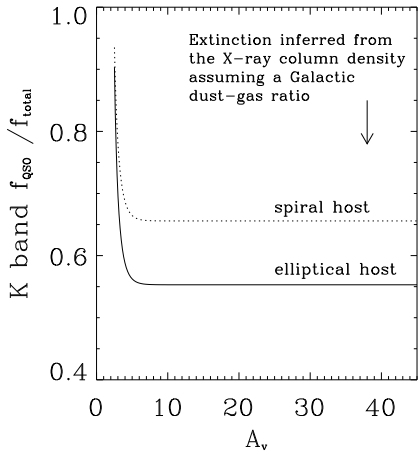}}
\end{picture}
\caption{The fractional AGN flux contribution to the total K
band flux required to yield the observed B--K colour for different absorbing
column densities. Elliptical galaxy; solid line. Spiral galaxy; dashed line.}  
\label{fig:absorp}
\end{figure}

The observed IR  colours of R--K=5.0, J--H=1.5 and
H--K=1.4 are all too red for a spiral or elliptical galaxy at $z=0.176$, where
one would expect  R--K=3.1, \citep{roche02}, and J--H=0.8, H--K=0.5
\citep{fioc99} suggesting that the dust-reddened nucleus is visible in the near--IR. 

We have estimated the relative contribution to the K band flux from
the active nucleus as a function of intrinsic reddening for both elliptical
and spiral host galaxy morphologies. This estimate is based on the B--K
colour, which has the largest available spectral leverage.
We used the elliptical and Sa spiral galaxy templates of \citet{manucci01}
and the QSO composite spectrum of \citet{francis91} extended to the K band.
For a range of extinctions we reddened the QSO composite spectrum 
assuming that any obscuring matter in the AGN has a Galactic
extinction curve \citep{fitz99} and determined the relative AGN and galaxy
contributions to the total  flux required to produce the observed B--K colour.

Figure \ref{fig:absorp} shows the allowed combinations of nuclear 
contribution versus intrinsic reddening for an elliptical and for a spiral
host. For an elliptical host the relative contribution from the active nucleus
is always greater than 55 percent and rises sharply with decreasing absorption.
For a spiral host the minimum active nucleus contribution is 66 percent.
Regardless of host galaxy morphology or the degree of reddening the AGN 
makes a substantial contribution to the near--IR flux.

\begin{figure}
\setlength{\unitlength}{1in}
\begin{picture}(3.5,3.3)
\put(-0.1,0.0){\includegraphics{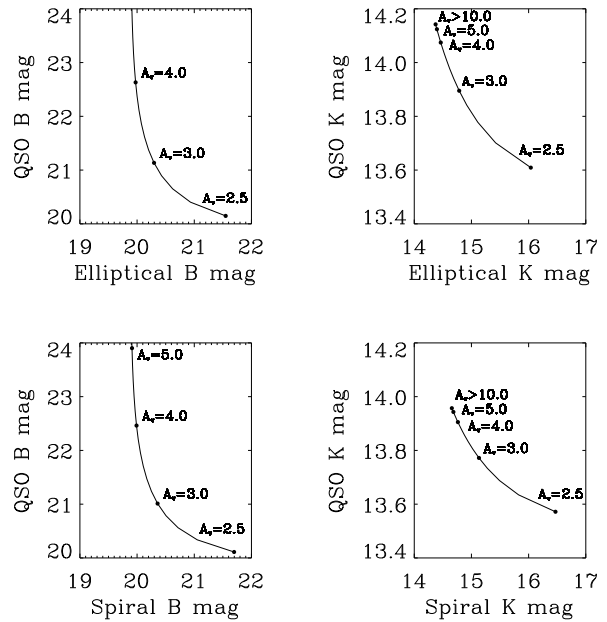}}
\end{picture}
\caption{The host galaxy and AGN magnitudes in the B and K bands
allowed in order to reproduce the observed B--K colour. }  
\label{fig:host}
\end{figure}

We have also explicitly calculated the B band and K band magnitudes for the
host and active nucleus corresponding to the permitted range of reddening and
flux contributions shown in Figure \ref{fig:absorp}.  
The results are illustrated in Figure \ref{fig:host}.
As the left panel shows, unless the AGN is reddened by less than 
$A_{v}\sim5$ (equivalent to $N_H\sim10^{22} {\rm~cm^{-2}}$), 
the host galaxy is responsible for virtually all the B band flux.

\section{Optical spectrum}

The measured [OIII]/${\rm H_\beta}$ ratio for \saxx\ is $\sim10$, of order 
$20\times$ larger than typically observed in
optically selected QSOs. This suggests that \saxx\ is extinguished by at least
$A_{v}=3.3$, 
corresponding to a minimum equivalent neutral hydrogen column
density of $6.5\times10^{21}\rm{~cm^{-2}}$. 

Assuming a galaxy flux contribution of 14 percent in the R band at $z=0$ for
typical unabsorbed AGN \citep{hutch84}, we have matched
the observed equivalent width of the broad $H_\beta$ component in our optical
spectrum with that of an obscured QSO plus host galaxy model spectrum.
We find an optical extinction of  $A_{v}\ge 4.0$, similar
to that derived via the observed [OIII]/${\rm H_\beta}$ ratio.
Comparison with our X--ray determined column density of $8.2\times10^{22}
\rm{~cm^{-2}}$ implies that the
ratio of optical/IR extinction to X--ray absorption  in \saxx\ is at least $\frac{1}{10}$ Galactic.

Another estimate of the optical extinction  may be derived via the observed
optical to X--ray slope, \aox\footnote{defined as the ratio of the UV flux
at 2500\AA\ to that of the X--ray flux at 2keV}. 
About 90\% of low redshift unobscured QSOs have $1.2< \alpha_{\rm ox} <1.7$
\citep{laor97}. Starting from the unabsorbed 0.5--2.0 keV
flux ($1.1\times10^{-12}\rm{~erg~cm^{-2}~s^{-1}}$) we determined the spread of
2500\AA\ flux consistent with this range of \aox, and converted these flux
values to B and K band magnitudes using the template of \citet{francis91}. 
Taking into account the uncertainty on the measured X--ray flux, 
we expect unreddened B and K magnitudes of $15.3^{+3.3}_{-1.9}$ and 
$12.3^{+3.3}_{-1.9}$ respectively.
Assuming an AGN contribution of between 50--100 percent in the K band, the 
observed K band magnitude of 13.5 implies optical extinction in the
range $A_{v}\le 29$.
Combining this upper limit  with the lower limit derived from
$H_\beta$  we conclude that the ratio of optical/IR extinction to X--ray
absorption in \saxx\ is $0.1-0.7$ Galactic.

\section{Discussion}

\subsection{Dust-gas ratio}

A low dust-gas ratio has been observed in several samples of obscured
AGN \citep[e.g.][]{maccacaro82,akiyama00,page01}.
This has been attributed to either a reduced dust content in these AGN, or as a
consequence of possible different emission regions, with X--ray emission
associated with deeper regions than optical emission \citep{reichert85}. 
\citet{maio01a} have also found dust-gas ratios significantly lower
than the standard Galactic value, 
for AGN with intrinsic luminosities $>10^{42}{\rm  ~erg~s^{-1}}$. 
However, they attribute this to a dust grain distribution biased in favour 
of large grains which flattens the extinction curve \citep{laordraine93}
and explains the reduced $A_{v}/N_{H}$ and  $E(B-V)/N_{H}$  
\citep{maio01b}. 
\citet{weinmurray2002} propose that most of the X--ray absorption in the
\citet{maio01a} galaxies occurs in a
region distinct from the region responsible for the optical extinction.

In \saxx\ the ratio of optical/IR extinction to X--ray absorption
is only $0.1-0.7$ times that expected assuming a Galactic dust-gas ratio, in
line with these previous studies. 
As this appears to be a widespread property of absorbed AGN it has
important consequences for the optical/IR appearance of the faint X--ray 
population.

\subsection{Implications for the faint source population}

Faint X--ray sources with 2--10 keV fluxes of $\sim10^{-14}{\rm ~erg~s^{-1}}$
are responsible for a large fraction of the X--ray background.
At these fluxes a high proportion of the sources are absorbed, with apparent
column densities of  $\sim10^{22}~{\rm cm^{-2}}$ \citep{rosati2002}. 
In many cases their optical counterparts are extremely red
\citep{barger01,alex02,crawford02}, confirming previous 
\rosat\ findings \citep{lehmann2000,lehmann01}.
However, understanding the nature of these sources has proved 
difficult due to their optical faintness. 
 
For those objects which are too faint for optical spectroscopy, 
photometric redshifts are estimated assuming  that the optical and IR 
emission is dominated by the host galaxy
\citep{alex02,lehmann01,franceschini01,crawford02}. 
The justification for this assumption comes from the observation that similar
sources for which optical spectroscopy is possible are dominated in the
optical by their host galaxies \citep{lehmann01}, 
and that the optical to IR colours of the faint
sources are consistent with those of elliptical or spiral galaxies
\citep{mainieri02,koekemoer02}. 

However, in \saxx\  a 
dust reddened AGN contributes at least 50 percent of the flux in the K band.
\saxx\ is similar both in its X--ray spectral properties and
in its optical colours to the sources which populate faint X--ray surveys although it is $\sim100\times$ brighter.
At $10^{-14}\rm{~erg~s^{-1}~cm^{-2}}$ we expect objects similar to \saxx\
to look identical to typical deep 
field sources and to be about 250 times more numerous on the sky than at a flux
of $8.6\times10^{-13}\rm{~erg~s^{-1}~cm^{-2}}$ \citep[assuming the type 1
luminosity function slope from][]{pagelf}. It is therefore possible that a
non-negligible fraction of faint sources could in fact have a significant 
AGN contribution in the near-IR.

Further evidence that significant numbers of faint absorbed sources have large
AGN contributions in the near--IR comes from the recent studies of
\citet{gandhi02} and \citet{stevens03}.
\citet{stevens03} found that three out of five of their X--ray selected EROs 
are point-like and therefore AGN dominated in the K band, while
\citet{gandhi02} found three out of eight of their sample of hard \chandra\
sources are point-like and have broad emission lines (i.e. are AGN dominated)
in the near--IR.

One significant consequence of the  
presence of a reddened AGN in the near--IR is that photometric redshift
determinations assuming a galaxy spectrum could be inaccurate. Weak, low
redshift objects could be assigned high redshifts in order to reproduce the red
colours from galaxy templates. An additional AGN contribution to the rest frame
near--IR flux should therefore be considered when determining photometric 
redshifts for sources with similar X--ray column densities to \saxx. 

\section{Conclusions}

We have presented the hard X--ray and optical spectra of the obscured QSO
\saxx. 
We find that the position reported for the optical counterpart
\citep[e.g.][]{fiore99} was incorrect. 
However, the redshift and optical spectrum given by \citep{fiore99} are correct.
The optical counterpart is an extremely red object with R--K=5.0 and
there is no \rosat\ counterpart. 

The X--ray spectrum is consistent with a Compton thin  absorbed power-law of 
photon index $2.0\pm0.2$  and intrinsic column density of
$N_{H}=8.2^{+1.1}_{-0.7}\times10^{22} {\rm~cm^{-2}}$.
The intrinsic 2--10 keV luminosity is
$1.1\pm0.1\times 10^{44}\rm{~erg~s^{-1}}$ which places
\saxx\ in the quasar regime. 
The  X--ray spectrum shows no evidence 
for a scattered component or for a soft component.

The colours indicate that the central AGN in
\saxx\ accounts for at least 50$\%$ of the K band emission regardless of
the assumed host morphology. This has important implications for photometric
redshift determinations for faint sources which assume that emission is 
dominated by a host galaxy.

Comparing X--ray and optically derived values for the column density intrinsic
to \saxx, we deduce that the ratio of extinction to X--ray absorption in \saxx\
is $0.1-0.7$ times that expected from a Galactic extinction curve.

\section*{Acknowledgements}

Based on observations obtained with \xmm, an ESA science mission
with instruments and contributions funded by ESA Member States and the USA
(NASA).
This publication makes use of data products from the Two Micron All Sky Survey,
which is a joint project of the University of Massachusetts and the Infra-red
Processing and Analysis Center/California Institute of Technology, funded by 
the National Aeronautics and Space Administration and the National Science 
Foundation.
This research made use of ALFOSC, which
is owned by the Instituto de Astrofisica  de Andalucia (IAA) and operated at
the Nordic Optical Telescope under agreement between IAA and the NBIfAFG of the
Astronomical Observatory of Copenhagen.

\end{document}